\documentclass[12pt]{article} 
\usepackage{epsfig} 
 1 
\def\dd{\partial}

\def\l{\lambda}

\def\({\left(}
\def\){\right)}

\def\citenum#1{{\def\@cite##1##2{##1}\cite{#1}}}
\def\citea#1{\@cite{#1}{}}

\def\l1vt{\vec{l_{1\perp}}}

\def\jol1{$J_0(\,l_{1\perp}\,r_{\perp}\,)$}

\def\beq{\begin{equation}}
\def\eeq{\end{equation}}
\def\bea{\begin{eqnarray}}
\def\eea{\end{eqnarray}}
\def\eq#1{{\mbox{Eq.\hspace{1mm}(\ref{#1})}}}

\def\scrbox#1{\mbox{\scriptsize #1}}

\def\npb#1#2#3{    {\it Nucl.\ Phys.\ }{\bf B#1} (#2) #3}

\def\plb#1#2#3{    {\it Phys.\ Lett.\ }{\bf B#1} (#2) #3}
\def\prd#1#2#3{    {\it Phys.\ Rev.\ }{\bf D#1} (#2) #3}

\def\prl#1#2#3{    {\it Phys.\ Rev.\ Lett.\ }{\bf #1} (#2) #3}

\def\zpc#1#2#3{    {\it Z.\ Phys.\ }{\bf C#1} (#2) #3}

\def\epj#1#2#3{    {\it Eur.\ Phys.\ J.\ }       {\bf #1} (#2) #3}

\newcommand{\email}[1]{${\!}^{\scrbox{#1)}}$}

\def\df2dlnq2{\dd{F_2}/\dd\log{Q^2}}

\relax
\begin{document}
\begin{titlepage}
\noindent

\vspace{1cm}
\begin{center}
  {\Large \bf LRG Production of Di-Jets As a} 
  \\[1.5ex]
  {\Large \bf Probe of s-Channel Unitarity}
  \\[4ex]

{\large \bf  Uri MAOR \email{1}}
\footnotetext{\email{1} Email: maor@post.tau.ac.il .}
\\[4.5ex]
{\it  School of Physics and Astronomy}\\
{\it  Raymond and Beverly Sackler Faculty of Exact Science}\\
{\it  Tel Aviv University, Tel Aviv, 69978, ISRAEL}\\[4.5ex]

\vspace{1cm}
{\bf THIS PAPER IS DEDICATED TO PROF. ROBERTO SALMERON}\\ 
{\bf ON HIS 80th BIRTHDAY}

\end{center}

\vspace{1cm}

{\samepage {\large \bf Abstract:}} 
Hard diffractive production of di-jets 
in the Tevatron and HERA are explored.
It is argued that a consistent pQCD description  
of these data can be attained 
after s-channel unitarity re-scattering 
screening corrections are 
incorporated in the calculation. 
To this end a simple procedure is presented
in which information derived from
the measured soft total, elastic and diffractive
cross sections simplifies the calculation and 
makes it parameter free.
The above approach eliminates the inconsistencies seemingly 
observed when comparing the rates of diffractive   
di-jets production in different channels.

\end{titlepage}
\section{Introduction}
\par
A large rapidity gap (LRG) in an hadronic or DIS 
final state is experimentally defined as a large gap in the 
$\eta-\phi$ lego plot where no hadrons are produced. A schematic 
description of a LRG di-jet diffractive final state is shown in Fig. 1.
Historically, LRG events where singled 
out\cite{Dok}\cite{Bj} as a 
signature for Higgs production 
due to a $W-W$ fusion sub process 
in hadron-hadron collisions at exceedingly high energies. 
At presently available energies, LRG di-jet central production 
is observed in the Tevatron\cite{D0}\cite{CDF}\cite{Dino} 
and HERA\cite{HERA}.
These measurements provide a unique opportunity to assess the asymptotic 
short distance behavior of the hard diffractive sub process 
responsible for the LRG di-jet in the final state and check our 
ability to adequately calculate it 
utilizing perturbative QCD (pQCD) methods. 
\begin{figure}
\begin{center}
\psfig{file=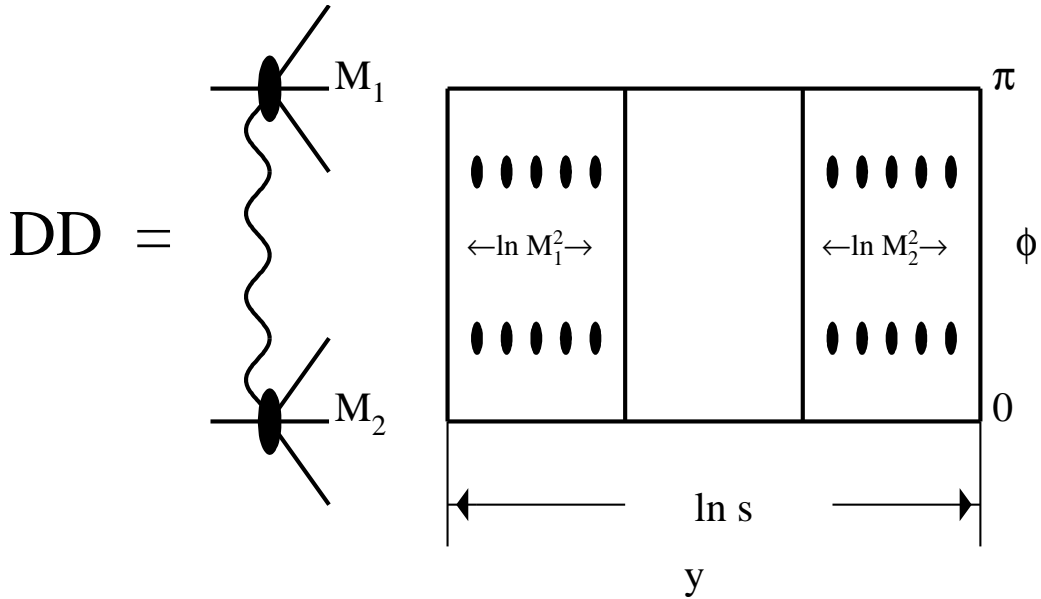,width=14cm,bbllx=133,bblly=466,bburx=492,bbury=680}
\end{center}
\caption{\it Lego plot for double diffractive dissociation.}
\end{figure}
\par
The production of two hard jets separated by a LRG can be 
explained by either:
\newline 
1) A fluctuation in the rapidity distribution of an inelastic event. 
The probability for such a fluctuation is proportional to 
$e^{-\Delta y/L}$, where $\Delta y=y_1-y_2$ and $L$ denotes the value of 
the correlation length. In a LRG event $\Delta y\gg L$ and, thus, the 
probability for such a fluctuation is very small. Or:
\newline
2) A QCD colorless t-channel exchange. We denote this 
exchange as a "hard Pomeron". Accordingly, we shall be interested in 
$F_s$, the ratio between the LRG and the inclusive 
cross sections for di-jet production.
It was noted by Bjorken\cite{Bj}
that we have to distinguish between 
the calculated pQCD ratio $F_s$ and the actual measured 
ratio $f_{gap}$, where the proportionality factor is called the 
"survival probability of a LRG", 
\begin{equation}\label{1.1} 
f_{gap}\,=\,\langle{\mid S \mid}^2\rangle \cdot F_s \,.
\end{equation}
\begin{figure}
\begin{center}
\psfig{file=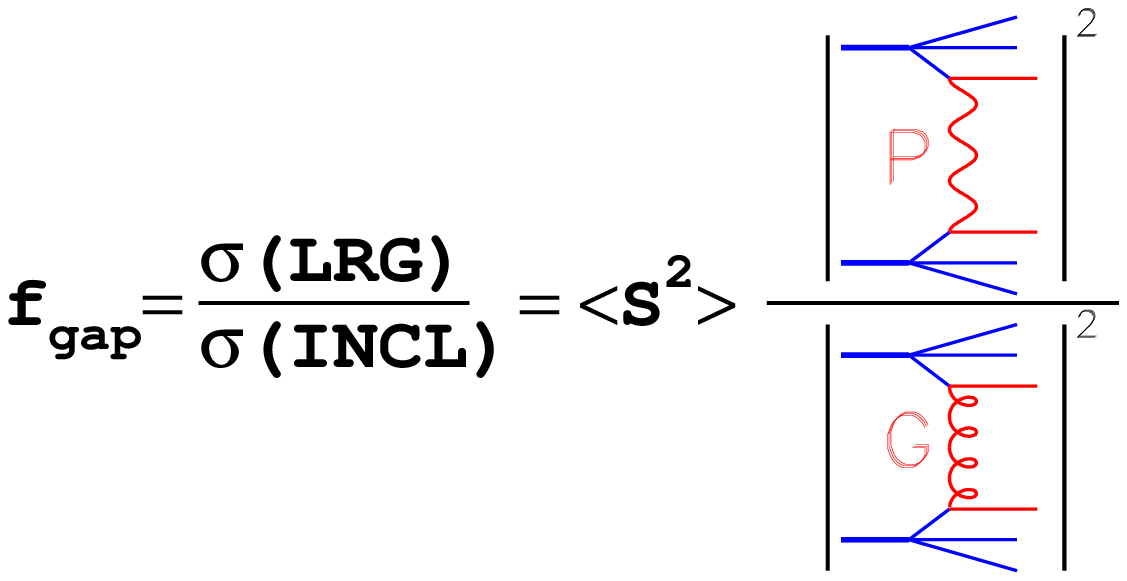,width=14cm,bbllx=150,bblly=530,bburx=480,bbury=720}  
\end{center}
\caption{\it Pictorial definition of $f_{gap}$.
P and G represent color singlet and octet respectively.}
\end{figure}
There have been several suggestions how to calculate the survival 
probability.
\newline
1) A simple procedure\cite{GLMLRG} has been suggested by Gotsman, Levin 
and Maor (GLM) 
to calculate $\langle{\mid S \mid}^2\rangle$. 
The procedure is based on the GLM model\cite{GLM} which incorporates 
the effects of s-channel unitarity in high energy soft scattering.
In the following I shall present this approach in some detail.
\newline
2) An alternative procedure has been suggested by the Durham 
group\cite{Durham}. It will be discussed briefly in Sec. 6.
\newline
3) Another alternative proposal\cite{Halzen}, 
associates the LRG phenomena with the color evaporation model (CEM).
Like the previous listed options it explains the observed low values of 
the survival probability using the interplay between hard and soft physics. 
I shall discuss it very briefly in Sec. 6.
\par
The survival probability is the probability of a given LRG not to be 
filled by debris (partons and/or hadrons). These debris can 
originate from the soft rescattering of the 
spectator partons, and/or from the gluon radiation emitted by partons
taking part in the hard interaction 
\begin{equation}\label{1.2}
\langle{\mid S(s,\Delta y) \mid}^2\rangle\,=\,
\langle{\mid S_{brem}(\Delta y) \mid}^2\rangle \cdot 
\langle{\mid S_{spec}(s) \mid}^2\rangle.
\end{equation}
$\langle{\mid S_{brem} \mid}^2\rangle$ can be calculated in pQCD\cite{S2}. Its  
estimate\cite{Durham}, at presently available energies,  is   
considerably smaller than $\langle\,{\mid S_{spec} \mid}^2\rangle$.
\par
The need for a reliable calculation of the survival probability is 
evident once we note that the values of $f_{gap}$, measured at the 
Tevatron\cite{D0}\cite{CDF} and HERA\cite{HERA}, are very small and 
energy dependent:
\begin{equation}\label{data1}
f_{gap}(\sqrt{s}=\,\,630\,GeV)\,=\,1.60 \pm 0.20\%\, (D0),\, 
and\,\, 2.70 \pm 0.90\%\,\, (CDF),
\end{equation}
\begin{equation}\label{data2}
f_{gap}(\sqrt{s}=1800\,GeV)\,=\,0.60 \pm 0.20\%\, (D0),\, 
and\,\, 1.13 \pm 0.16\%\,\, (CDF),
\end{equation}
\begin{equation}\label{data3}
f_{gap}(\sqrt{s}=\,\,150\,GeV)\,=\,7.00 \pm 2.00\%\,\, (ZEUS).
\end{equation}
These numbers are considerably smaller, and reduce with increasing energy, 
in contradiction to the pQCD expectations. These observations can 
be understood only if $\langle{\mid S \mid}^2\rangle$ 
is rather small and energy 
dependent. Indeed, a direct measurement of the energy dependence of 
$f_{gap}$ by D0\cite{D0} gives 
\begin{equation}\label{fratio}
\frac{f_{gap}(\sqrt{s}=\,\,630\,GeV)}{f_{gap}(\sqrt{s}=1800\,GeV)}\,=
\,3.4 \pm 1.3,
\end{equation}
to be compared with a ratio of $2.7 \pm 0.6$ obtained from the 
individual numbers just quoted (\eq{data1} and \eq{data2}).
As stated, the
purpose of this short review is to present a relatively simple, 
and parameter free, procedure to calculate 
the survival probability and correlate its small value with the properties  
of high energy soft scattering.
I shall show that the Tevatron data is 
perfectly consistent with the HERA data and that the calculations yield 
a natural interpretation of LRG di-jets observed in different channels. 
An important consequence of this analysis is that it 
provides significant support for the need to implement s-channel
unitarity corrections to soft hadronic interactions in the 
HERA - Tevatron energy range. 
\par
The organization of this paper is as follows: In Sec. 2 I briefly review  
the roll of s-channel unitarity in high energy soft scattering and its 
implications. This is followed in Sec. 3 by a suggested model by 
GLM\cite{GLM}, compatible 
with unitarity constraints, which is based on the eikonal approximation. 
The calculation of the survival probabilities in the eikonal 
approach are presented and discussed in Sec. 4. The procedure presented 
incorporates the soft scattering basic measurements which replaces 
the need for free parameters. A generalization to a multi channel 
rescattering model is briefly reviewed in Sec. 5.
A short discussion is presented in Sec. 6. 

\section{s-Channel Unitarity in High Energy Soft Scattering} 
\par
High energy soft scattering is commonly described by the 
Regge-pole model\cite{Regge}. The theory was introduced more than 
40 years ago and was soon after followed by a very rich 
phenomenology. The two key ingredients of this approach 
are the Pomeron whose t dependent trajectory is given by
\begin{equation}\label{2.1}
\alpha_P(t)\,=\,\alpha_P(0)\,+\,\alpha_P^{\prime} t,
\end{equation}
and a leading Regge trajectory
\begin{equation}\label{2.2}
\alpha_R(t)\,=\,\alpha_R(0)\,+\,\alpha_R^{\prime} t.
\end{equation}
Donnachie and Landshoff (DL) have vigorously promoted\cite{DL} an 
appealing and very simple Regge parametrization for total and 
forward differential elastic hadron-hadron cross sections in which 
\begin{equation}\label{2.3}
\sigma_{tot}\,=\,X\(\frac{s}{s_0}\)^{\epsilon}\,
+\,Y\(\frac{s}{s_0}\)^{-\eta},
\end{equation}
and
\begin{equation}\label{2.4}
\frac{d\sigma_{el}}{dt}\,=\,\(\frac{d\sigma_{el}}{dt}\)_0 e^{B_{el}t}.
\end{equation}
$\(\frac{d\sigma_{el}}{dt}\)_0$ and $\sigma_{tot}$ are related 
through the optical theorem. The forward elastic high energy exponential 
slope is given by 
\begin{equation}\label{2.4.1}
B_{el}\,=\,B_0\,+\,2\alpha_P^{\prime}ln\(\frac{s}{s_0}\).
\end{equation}
The data is excellently fitted with universal parameters: 
\newline
1) A supercritical Pomeron with an intercept 
$\alpha_P(0)\,=\,1\,+\,\epsilon$, where $\epsilon\,=\,0.08$ 
accounts for 
the high energy growing total cross sections. Its fitted\cite{BKW} slope 
value is $\alpha_P^{\prime}\,=\,0.25\,GeV^{-2}$. 
\newline
2) The low energy data is controlled by the leading Regge trajectory 
whose universal parameters are\cite{DL} $\eta\,=\,0.45$ 
and $\alpha_R^{\prime}\,=\,1\,GeV^{-2}$.
\newline
DL offer a global fit to all available hadron-hadron and photon-hadron 
total and elastic cross section data. Note, though, that in reality only 
$\bar p p$ and $\gamma p$ reactions have attained high enough energies in 
which the Pomeron parameters can be tested. We also note that the 
comprehensive DL analysis is limited to the properties of the elastic 
amplitude with no reference to diffractive scattering.
\par
Elastic and diffraction dissociation are similar processes which have 
predominantly forward imaginary amplitudes corresponding to the exchange 
of vacuum quantum numbers in the t-channel. As such, both channels are 
dominated in the high energy limit by a Pomeron exchange and are expected, 
in a simple Regge model, to exhibit rather similar dependences on the 
kinematic variable. Indeed, in the triple Regge limit applied to 
high energy single diffraction (SD) we get\cite{Mueller}
\begin{equation}\label{2.5}
\frac{M^2d\sigma_{sd}}{dM^2dt}\,=
\,\sigma^2 (s_0)\(\frac{s}{M^2}\)^{2\epsilon+2\alpha_P^{\prime} t}
G_{PPP}\(\frac{M^2}{s}\)^{\epsilon},
\end{equation}
where $M$ is the diffracted mass, and $G_{PPP}$ is the triple Pomeron 
vertex coupling. The virtue of this 
formalism is that it provides a strong correlation between the 
energy dependences of $\sigma_{tot}, \sigma_{el}, B_{el}$, 
and the energy and $M^2$
dependences of $\sigma_{sd}$.
\par
The simple model just presented is bound, eventually, to violate 
s-channel unitarity since $\sigma_{el}$ and $\sigma_{sd}$ grow with energy 
as $s^{2\epsilon}$, modulu logarithmic corrections, while $\sigma_{tot}$
grows only as $s^{\epsilon}$. 
The detailed theoretical problems at stake are easily identified 
in an impact b-space formalism which is outlined below.
\par
The elastic scattering amplitude is normalized so that
\begin{equation}\label{2.6}
\frac{d\sigma_{el}}{dt}\,=\,\pi\mid f_{el}(s,t) \mid ^2,
\end{equation}
and
\begin{equation}\label{2.7}
\sigma_{tot}\,=\,4 \pi Im f_{el}(s,0).
\end{equation}
The elastic amplitude in b-space is defined as
\begin{equation}\label{2.8}
a_{el}(s,b)\,=\,\frac{1}{2\pi}\int d{\bf q} e^{-i{\bf q\cdot b}} 
f_{el}(s,t),
\end{equation}
where $t\,=\,-q^2$.
In this representation
\begin{equation}\label{2.9}
\sigma_{tot}\,=\,2\int d{\bf b} Im a_{el}(s,b),
\end{equation}
\begin{equation}\label{2.10}
\sigma_{el}\,=\,\int d{\bf b} \mid a_{el}(s,b) \mid ^2.
\end{equation}
\par
The common formulation of s-channel unitarity implies that 
$\mid a_{el}(s,b)\mid\leq 1$, which is the black bound. 
We also have the analyticity/crossing bound. The position of this bound 
depends on the mass of the lightest exchange in the t-channel. 
Froissart\cite{Froissart}, at the time, considered it to be a $\pi$ meson.
Our present thinking tends to consider it as the lightest glue-ball 
which can be exchanged in the $t$ channel. 
The Froissart bound is obtained
when the unitarity and analyticity/crossing bounds are combined. 
\begin{equation}\label{2.11}
\sigma_{tot}\, \leq \, C \, ln \(\frac{s}{s_0}\),
\end{equation}
where $C$ depends on the crossed channel exchanged mass. 
Note that both the unitarity and 
Froissart bounds are numerical, not functional. Consequently, the 
Froissart bound (with any reasonable value of $C$) is not applicable to  
present day accelerator physics, 
being much higher than the actual data. 
\begin{figure}
\begin{center}
\psfig{file=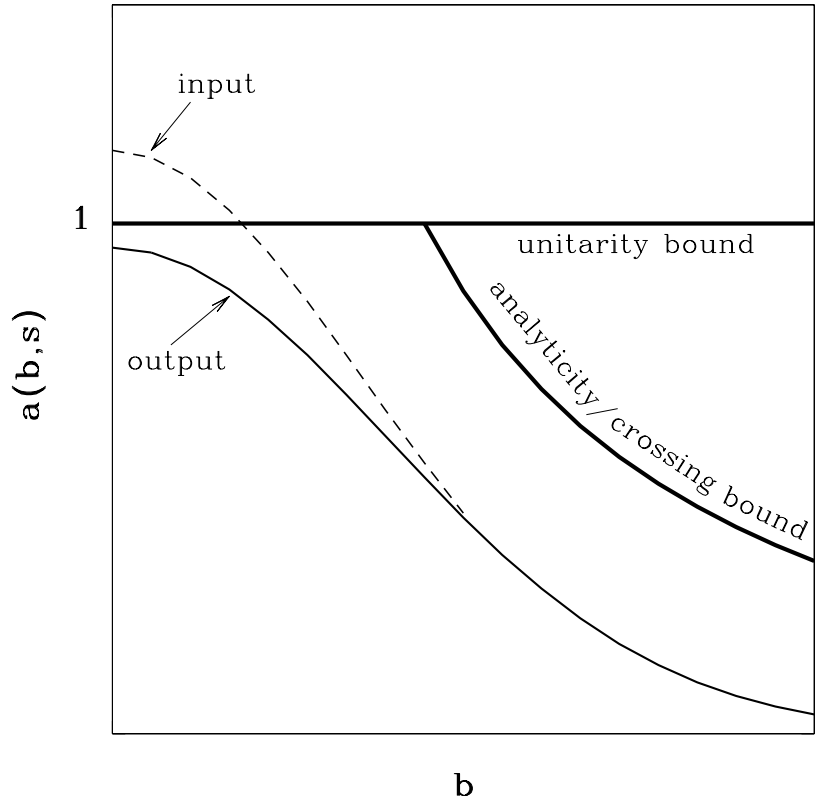,width=12cm}
\end{center}
\caption{\it A pictorial illustration of a high energy  
b-space elastic amplitude bounded by unitarity and analyticity/crossing.
In the illustration we have an input amplitude which violates the 
unitarity bound and an output amplitude obtained after a unitarization 
procedure.}
\end{figure}
On the other hand, as
we shall see, the unitarity bound is very relevant to the analysis of the 
presently available high energy data.
A schematic illustration of the above is presented
in Fig. 3.
\par
As noted, a simple Regge pole parametrization will eventually violate 
s-channel unitarity since the integrated elastic and diffractive cross 
sections grow with s much faster than the total cross section. The 
question is if this is a future problem to be confronted only at far 
higher  
energies than presently available, or is it a phenomena which can 
be identified through experimental signatures observed within the 
available high energy data base. 
It is an easy exercise to check that the DL 
model\cite{DL}, with its fitted global parameters, will violate the 
unitarity black limit for small b just above the present Tevatron energy. 
Indeed, CDF reports\cite{CDFb} that 
$a_{el}(b=0,\sqrt{s}=1800)\,=\,0.96\pm 0.04$. 
In a DL type model this should be reflected in $\epsilon$ acquiring 
an energy dependence rather than the fixed fitted value. This is not 
observed, as yet, in the total and elastic cross section available data. 
We note, though, that 
the energy dependence of the experimental SD cross section\cite{Dino}  
in the ISR-Tevatron energy range is
much weaker than the power dependences observed for 
$\sigma_{el}$. This observation is dramatically demonstrated once we 
check the ratio between $\(\frac{d\sigma_{sd}}{dt}\)_0$ and 
$\(\frac{d\sigma_{el}}{dt}\)_0$ which goes down with energy.
\section{The GLM Screening Correction Model}
\par
The theoretical difficulties, pointed out in the previous section,  
are eliminated once we take into account the corrections 
necessitated by unitarity. The problem is that enforcing unitarity 
is a model dependent procedure. In the following I shall confine myself 
to the GLM screening corrections (SC) model\cite{GLM}. 
In this semi realistic model the unitarity enforced SC are calculated 
utilizing  
a Glauber type eikonal approximation in which the scattering matrix is 
diagonal and only repeated elastic 
rescatterings are summed. Accordingly, we write
\begin{equation}\label{2.12}
a_{el}(s,b)\,=\,i\(1-e^{-\Omega(s,b)/2}\),
\end{equation}   
where the opacity $\Omega$ is a real function 
which is identified with the value of 
the imaginary part of a DL input super critical Pomeron. Analyticity 
and crossing symmetry are easily restored by the substitution 
$s^{\alpha_P}\,\rightarrow\,s^{\alpha_P}e^{-i\pi \alpha_P/2}$. 
Combining \eq{2.9}, \eq{2.10} and \eq{2.12} we get
\begin{equation}\label{2.12.1}
\sigma_{tot}\,=\,2 \int d^2 b \(1\,-\,e^{-\Omega(s,b)/2}\),
\end{equation}
and
\begin{equation}\label{2.12.2}
\sigma_{in}\,=\,2 \int d^2 b \(1\,-\,e^{-\Omega(s,b)}\).
\end{equation}
\par
Since the scattering matrix is diagonal, the unitarity constraint is 
written as
\begin{equation}\label{2.13}
2 Im a_{el}(s,b)\,=\,{\mid a_{el}(s,b) \mid}^2\,+\,G^{in}(s,b).
\end{equation}
Note that \eq{2.12} is a solution of \eq{2.13}. We also get
\begin{equation}\label{2.14}
G^{in}\,=\,1\,-\,e^{-\Omega(s,b)}.
\end{equation}  
The eikonal approximation can be summed analytically with a Gaussian input
which corresponds to an exponential representation of the scattering
amplitude in t-space (see \eq{2.4}). This is, clearly, an over
simplification, but it reproduces well the small t elastic cross section
which contains the bulk of the data.
\begin{equation}\label{2.15}
\Omega(s,b)\,=\,\sigma(s_0) \Gamma_S(s,b)\,=\,\nu(s) 
e^{\frac{-b^2}{R_S^2(s)}},
\end{equation}
where
\begin{equation}\label{2.16}
\nu(s)\,=\,\Omega(s,b=0)\,=\,
\frac{\sigma(s_0)}{\pi R_S^2(s)}\(\frac{s}{s_0}\)^{\Delta},
\end{equation}
and  
\begin{equation}\label{2.17}
R_S^2(s)\,=\,4R_0^2\,+\,4\alpha_P^{\prime} ln (\frac{s}{s_0}).
\end{equation}
The soft profile is defined
\begin{equation}\label{2.18.0}
\Gamma_S(s,b)\,=\,\frac{1}{\pi R_S^2(s)} e^{-\frac{b^2}{R_S^2(s)}}.
\end{equation}
Some clarifications
are in order so as to distinguish between 
the GLM model input and output:
\newline
1) The power $\Delta$, and accordingly $\nu$, are input information and
should not be confused with DL effective power $\epsilon$ and its 
corresponding cross section. Obviously, $\Delta\,>\,\epsilon$.
\newline
2) The relation
$B_{el}\,=\,\frac{1}{2}R_S^2(s)$ depends mostly on the high $b$ tail 
of the profile $\Gamma_S$. As such, SC do not change 
the output soft radius and this relation significantly.
\par
With this input we get
\begin{equation}\label{2.18}
\sigma_{tot}\,=\,2 \pi R_S^2(s)\( ln (\frac{\nu(s)}{2})\,+\,C\,
-\,Ei(-\frac{\nu(s)}{2}) \)
\,\propto\, ln^2(s),
\end{equation}
\begin{equation}\label{2.19}
\sigma_{in}\,=\,\pi R_S^2(s)\( ln [\nu(s)]\,+\,C\,-\,Ei[- \nu(s)] \)
\,\propto \, \frac{1}{2} ln^2(s),
\end{equation}
\begin{equation}\label{2.20}
\sigma_{el}\,=\,\sigma_{tot}\,-\,\sigma_{in}\,=\,
\pi R_S^2(s)\(ln (\frac{\nu(s)}{4})\,+\,C\,-\, 2 Ei(-\frac{\nu(s)}{2})\,
+\,Ei[-\nu(s)]\)\,
\propto \,\frac{1}{2} ln^2(s),
\end{equation}
where $Ei(x)\,=\,\int_{-\infty}^x \frac{e^t}{t} dt$, and $C\,=\,0.5773$ 
is the Euler constant.
\par
The formalism just presented can be extended\cite{GLM} also to diffractive 
channels. The key observation is traced to \eq{2.14} from where we define 
\begin{equation}\label{2.21.0}
P_S(s,b)\,=\,e^{-\Omega(s,b)}
\end{equation}
to be the probability that the two initial hadrons do not interact 
inelastically. i.e.
there is no initial state rescattering in the 
inelastic sector at the given $s,b$ values.  
Accordingly, the screened diffractive cross section is obtained 
by multiplying its b-space amplitude by this probability.
In this approximation rescatterings in the 
diffractive final state are not included. 
Specifically, for SD we take the 
b-space transform of \eq{2.5} multiplied by $P_S(s,b)$
and obtain after integration
\begin{equation}\label{2.21}
\frac{M^2 d\sigma_{sd}}{d M^2}\,=\,
\frac{\sigma^2(s_0)}{2\pi \bar R_S^2(\frac{s}{M^2})}
\(\frac{s}{M^2}\)^{2\Delta} G_{PPP} \(\frac{M^2}{s_0}\)^{\Delta}
a_{D}(s,M^2) \frac{1}{[2\nu(s)]^{a_{D}(s,M^2)}} 
\gamma[a_{D}(s,M^2),2\nu(s)],
\end{equation}
where
\begin{equation}\label{2.22}
\bar R_S^2\(\frac{s}{M^2}\)\,=\,2R_0^2\,+\,r_0^2\,
+\,4\alpha_P^{\prime} ln\(\frac{s}{M^2}\).
\end{equation}
$r_0 \leq 1 GeV^{-2}$ denotes the radius of the triple Pomeron vertex and 
can be safely neglected.
Note that $\frac{1}{2}R_S^2\,<\,{\bar R}_{S}^2\,<\,R_S^2$. In the 
ISR-Tevatron energy range ${\bar R}_S^2$ is just moderately larger than 
$\frac{1}{2}R_S^2$.
\begin{equation}\label{2.23}
a_{D}(s,M^2)\,=\,\frac{2 R_S^2(s)}{\bar R_S^2(\frac{s}{M^2})\,
+\,2 \bar R_S^2(\frac{M^2}{s_0})}.
\end{equation}
$\gamma(a,2\nu)$ denotes the incomplete Euler gamma function 
\begin{equation}\label{2.24.0}
\gamma(a,x)\,=\,\int_0^x\,z^{a-1}e^{-z}dz.
\end{equation}
In the high energy limit \eq{2.21} simplifies to
\begin{equation}\label{2.24}
\frac{M^2 d\sigma_{sd}}{dM^2}\,
=\,\pi R_S^2(s) G_{PPP} \(\frac{M^2}{s_0}\)^{\Delta}\,\propto ln(s).
\end{equation}
\par
The simple model, just presented, provides a significant improvement on  
the Regge DL model which serves as its input:
\newline
1) It is compatible with unitarity (and Froissart) behavior in the 
high energy limit. 
In particular, the asymptotic ratio of 
$\frac{\sigma_{el}}{\sigma_{tot}}$ is kept at $\frac{1}{2}$ 
(\eq{2.18} - \eq{2.20}).
\newline
2) Even though both $\sigma_{tot}$ and $\sigma_{el}$ change, at high 
enough energies, their dependence on $s$ from a power to a 
$ln^2(s)$ behavior, this change will be experimentally observed at an 
energy far higher than presently available. The model is, thus, compatible 
with DL in the ISR-Tevatron energy range. 
\newline
3) The SD cross section changes its power dependence on $s$ to a $ln(s)$ 
behavior. Checking the numerics of \eq{2.24} this change is significant 
in the ISR-Tevatron energy range. 
Note that in the high energy limit $\frac{\sigma_{diff}}{\sigma_{tot}}$ 
vanishes like $\frac{1}{ln (s)}$.
\newline
4) The significant difference between the energy dependence of 
the seemingly similar elastic and SD (and other diffractive)  
cross sections is understood once we 
examine their b-space amplitudes. The elastic amplitude is central 
(approximately given by a central Gaussian). Its DL parametrization implies 
that $a_{el}(s,b)$ will violate unitarity at small b somewhere between 2 
and 3 $TeV$ c.m. energy. Since this is confined to exceedingly  
small b values, it translates 
to very small changes in the normalization of the total and elastic cross 
sections. Screening is very different for the SD b-space amplitude. The 
screening suppression is obtained from the multiplication by 
$P_S(s,b)$, of \eq{2.21.0}, which is very small at small b and 
approaches unity at higher b values. Whereas the screened elastic b-space 
amplitude maintains its centrality with a maximum at $b\,=\,0$, the 
screened SD amplitude changes from a central to a 
peripheral b distribution, which is accompanied by a significant reduction 
of the output SD cross section.
This phenomena is compatible with the Pumplin bound\cite{Pumplin} in which 
the sum of the elastic and diffractive b-space amplitudes is bounded by 
$\frac{1}{2}$.
\par 
Regardless of its qualitative success, the toy GLM model does not 
provide a realistic reproduction of the data. Its two important 
deficiencies are:
\newline
1) The model does not produce a good reproduction of the SD 
data\cite{Dino}. This is amended in a multi channel 
eikonal model\cite{GLMLRG} 
in which diffractive intermediate rescatterings are also included.
\newline
2) The model is applicable only for small $t$ elastic scattering and it 
does not reproduce the high $t$ diffractive dip experimentally observed 
at high energies.
This is a consequence of our usage of a Gaussian b profile. To improve 
this deficiency we have to use a more elaborate profile\cite{BSW}.
\section{Survival Probability of LRG in the GLM Model} 
Following Bjorken\cite{Bj}, the survival probability, associated with the 
soft rescatterings of the spectator partons, is defined as the normalized 
integrated product of two quantities:
\begin{equation}\label{3.1}
\langle{\mid S_{spec} \mid}^2\rangle\,=\,
\frac{\int d^2 b \Gamma_H(s,b) P_S(s,b)}
{\int d^2 b \Gamma_H(s,b)}.
\end{equation}
1) The first element in \eq{3.1} is a convolution over the parton 
densities of the two interacting hadronic projectiles. 
This provides the uncorrected hard 
parton-parton collision cross section, leading, in our context, to 
di-jets.  
Following the discussion of soft scattering in Sec. 2, we assume a 
Gaussian hard profile 
\begin{equation}\label{3.2}
\Gamma_H\,=\,\frac{1}{\pi R_H^2(s)} e^{-\frac{b^2}{R_H^2(s)}},
\end{equation}
where $R_H$ denotes the radius of the hard scattering process. This choice 
enables an analytical solution of \eq{3.1}. More elaborate choices require 
a numerical evaluation of this equation.
\newline
2) The second element is the probability $P_S(s,b)$, defined in 
\eq{2.21.0}, that no inelastic 
soft interaction takes place between the incoming projectiles 
at impact parameter $b$ and c.m. energy square $s$. We define
\begin{equation}\label{3.3}
a_H(s)\,=\,\frac{R_S^2(s)}{R_H^2(s)}\,=\,\frac
{interaction\,area\,for\,soft\,collisions}
{interaction\,area\,for\,hard\,collisions}\,>\,1.
\end{equation}
$a_H(s)$ grows logarithmically with $s$.
As stated, \eq{3.1} can be analytically evaluated with our choice of 
Gaussian profiles and we get
\begin{equation}\label{3.4}
\langle{\mid S_{spec} \mid}^2\rangle\,=\,\frac{a_H(s) \gamma[a_H(s),\nu(s)]}
{[\nu(s)]^{a_H(s)}}.
\end{equation}
$\nu(s)$ was defined in \eq{2.16} and the $\gamma$ function in 
\eq{2.24.0}.
\par
A straight forward estimate of the LRG survival probabilities can be 
derived\cite{Bj}\cite{GLMLRG}\cite{Durham}\cite{Halzen} from various 
models assumed to describe the soft and hard interactions. 
An alternative method suggested by GLM\cite{GLMLRG} is based on the 
GLM eikonal model reviewed in Sec. 3. 
This calculation of $\langle{\mid S_{spec} \mid}^2\rangle$ 
is executed with no further recourse to the theoretical 
models, utilizing just the experimental data from which we can directly 
deduce the input needed to solve \eq{3.4}. 
To this end we have to assess the values of $R_H^2$, $R_S^2$ and $\nu(s)$.
\par
The value of $R_H^2$ is estimated from two independent experiments:
\newline
1) In analogy to the relation $R_S^2\,\simeq 2 B_{el}$, we can estimate 
the value of the hard radius from the exponential slope of a clean hard 
process $R_H^2\,\simeq 2 B_{hard}$. We use the excellent 
HERA data\cite{HERAJ} on $\gamma+p \rightarrow J/\Psi+p$, where 
$B_{J/\Psi}\,\simeq \,4 GeV^{-2}$ with a small logarithmic dependence on 
the incoming energy. 
\newline
2) The above is compatible with the CDF estimate of 
the double parton cross section. This cross section is connected through 
factorization to the effective hard cross section
\begin{equation}\label{3.5}
\sigma_{DP}\,=\,
m \frac{\sigma_{incl1}(2\,jets)\,\sigma_{incl2}(2\,jets)}
{2 \sigma_{eff}^H},
\end{equation}
where the factor $m$ is equal to 1 for identical pairs and 2 for
different pairs.
$\sigma_{eff}^H$ relates directly to
the hard profile (\eq{3.2})
\begin{equation}\label{3.6}
\frac{1}{\sigma_{eff}^H}\,=\,\frac{1}{2 \pi R_H^2}\,=\,
\int d^2 b \Gamma_H^2(b).
\end{equation}
The experimental value\cite{CDF2p} 
$\sigma_{eff}^H\,=\,14.5\pm1.7\pm2.3\,mb$ suggests that 
$R_H^2\,=\,5-7 GeV^{-2}$.
\newline
In our calculations we took the HERA deduced value $R_H^2\,=\,8 GeV^{-2}$.
This is a conservative choice which may be slightly changed with the 
improvement of the Tevatron estimates.
\par
The values of $R_S^2(s)$ are taken from the experimental elastic 
slope of the high energy $\bar p p$ data.
The values of $\nu(s)$ can be obtained in the GLM model directly from the 
measured values of $\frac{\sigma_{el}}{\sigma_{tot}}$. Checking \eq{2.18} 
and \eq{2.20}. We note that the (s dependent) ratio of elastic to total 
high energy cross sections provides explicit information on $\nu(s)$.
Once we have determined $\nu(s)$ and $a_H(s)$, the survival probability is 
obtained from \eq{3.4}. 
\par
The important result of our calculation, which is displayed in Fig. 4, is 
that $\langle{\mid S_{spec} \mid}^2\rangle$
\begin{figure}
\begin{center}\psfig{file=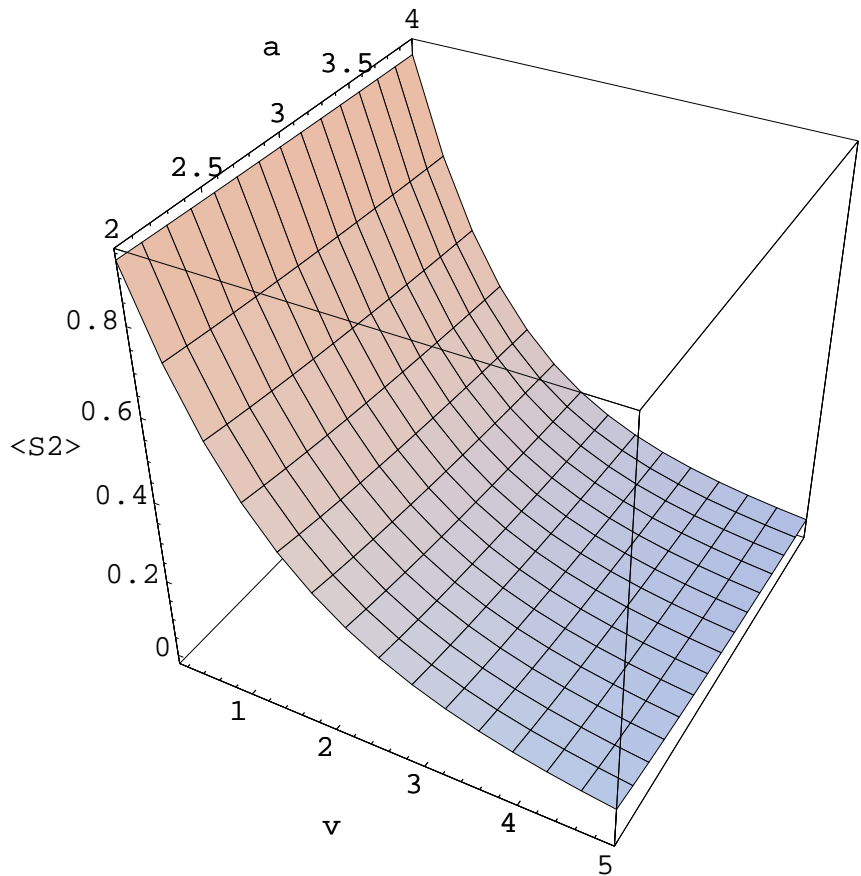,width=12cm, bbllx=100,bblly=290,bburx=520,bbury=706}
\end{center}
\caption{\it $\langle{\mid S_{spec} \mid}^2\rangle$
as a function of $\nu(s)$ and $a_H(s)$.}
\end{figure}
decreases as we go to higher energies. This is traced to the 
experimentally observed increase with energy  
of $\nu(s)$ and $a_H(s)$. 
To further illustrate this feature, we show a contour plot of
\begin{figure}
\begin{center}
\psfig{file=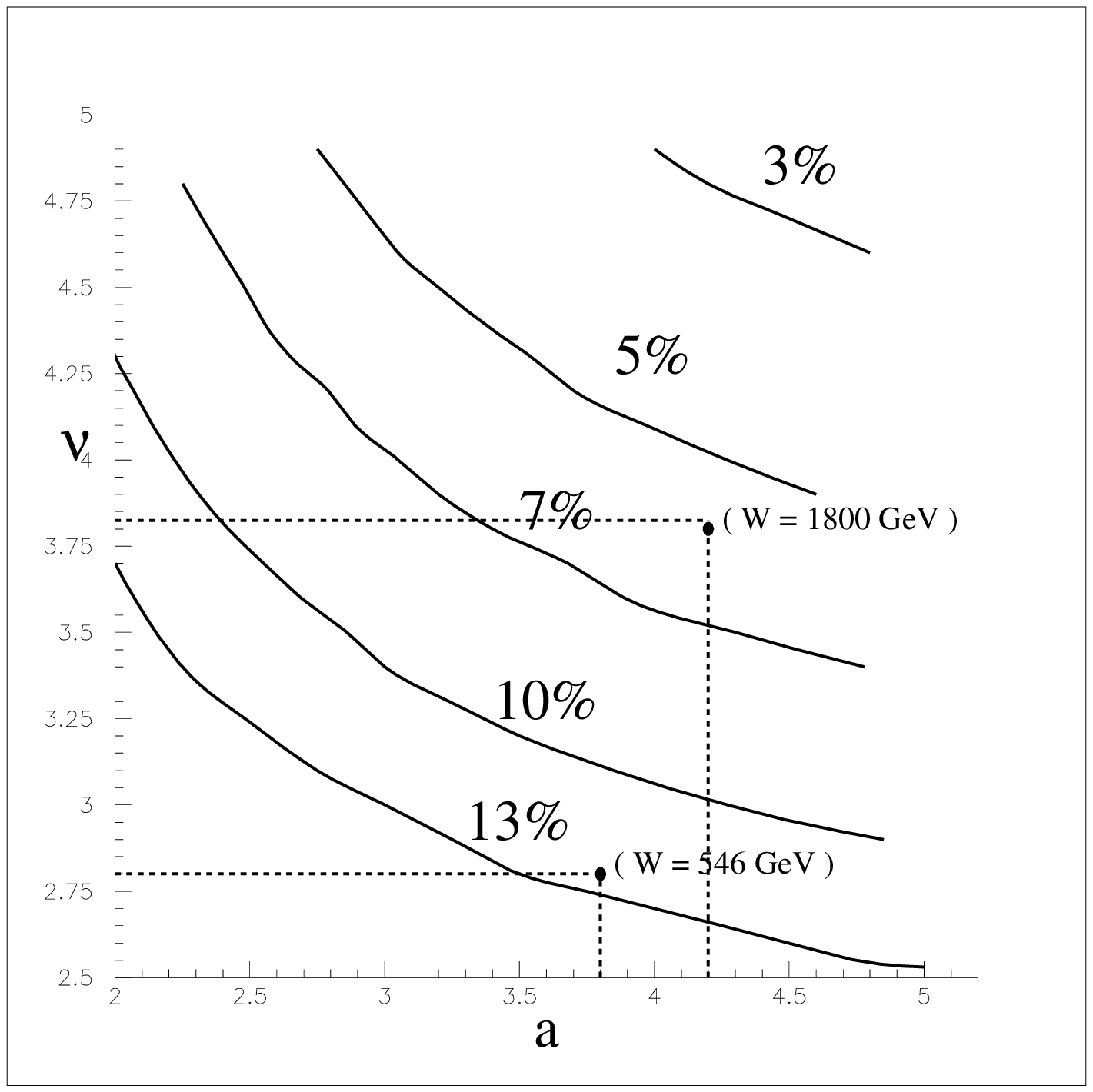,width=12cm,bbllx=100,bblly=300,bburx=510,bbury=710}
\end{center}
\caption{\it A contour plot of $\langle{\mid S_{spec} \mid}^2\rangle$
against $\nu(s)$ and $a_H(s)$.}
\end{figure}
$\langle{\mid S_{spec} \mid}^2\rangle$ against $\nu(s)$ and $a_H(s)$ in Fig. 5.
\par
{\samepage
Our results for the survival probability at the reported energies:
\begin{center}
\begin{tabular}{|c|c|}
\hline
\multicolumn{1}{|c|}{$\sqrt{s}\,\,$(GeV)}&
\multicolumn{1}{c|}{$\langle{\mid S_{spec} \mid}^2\rangle$}\\
\hline
 150 & 16.3\% \\
 630 & 12.8\% \\
1800 &  5.9\% \\
\hline
\end{tabular}
\end{center}
provide a qualitative explanation for the smallness and energy dependence 
observed experimentally.}
However, for a quantitative comparison between the reported $f_{gap}$ data 
(\eq{data1} - \eq{data3}) 
and our results we have to incorporate a pQCD estimate of $F_s$.   
I shall return to this issue in the discussion. To bypass this difficulty 
it is constructive to check the ratio of the survival probabilities 
calculated at 630 and 1800 $GeV$
\begin{equation}\label{GLMr}
\frac{{\langle{\mid S_{spec} \mid}^2\rangle }_{\sqrt{s}\,=\,630}}
{{\langle{\mid S_{spec} \mid}^2\rangle }_{\sqrt{s}\,=\,1800}}\,=\,2.2\pm 0.2.
\end{equation}
Our calculated ratio is fully substantiated by the D0 result\cite{D0} in 
which the corresponding survival probability 
(rather than the $f_{gap}$) ratio was estimated at $2.2 \pm 0.8$. 
Note that for our calculations at 630 $GeV$ we had to interpolate the 
input data information from 546 $GeV$. 
\section{Extension to a Multi Channel Model}
In as much as the GLM results on the survival probability are very 
satisfactory, the over all quality of the GLM predictions is 
not that good. The GLM model\cite{GLM} was originally conceived so 
as to explain the exceptional mild energy dependence of soft diffractive 
cross sections. It suggested that a high energy change from a 
power to a logarithmic energy  
dependence of the diffractive cross sections is a 
signature indicating the onsetting of 
s-channel unitarity corrections. 
The model actual calculations do, indeed, result in a qualitative 
significant difference between the energy dependence of the total 
and elastic cross sections, on the one hand, and diffractive cross 
sections, on the other hand. However, they do not provide an 
adequate reproduction of the SD cross sections\cite{Dino}
in the ISR-Tevatron energy range. A possible remedy to this 
deficiency of the GLM model is to replace the one channel 
eikonal model, in which only elastic rescatterings are included, 
with a multi channel eikonal model in which inelastic diffractive  
intermediate rescatterings 
are included as well. This is not a difficult technical 
problem\cite{GLMMC}, but, in our context, we have to secure that a 
multi channel model does improve the diffractive (specifically 
SD) predictions of the GLM model, 
while maintaining, simultaneously, its 
excellent results on LRG\cite{GLMLRG}. 
\par
The implicit assumptions of the simple GLM model are:
\newline
1) Hadrons are the correct degrees of freedom at high energies.
\newline
2) At high energy $Re\,a_{el}\,\ll\,Im\,a_{el}$.
\newline
3) $\sigma_{diff}\,\ll\,\sigma_{el}$.
\newline
4) Only the fastest partons can interact with each other.
\newline
Clearly, the last two assumptions are the weakest of the four.
There is no question that at small enough $x$ (high enough 
energy) one has to take into account 
$\langle{\mid S_{brem}(s) \mid}^2\rangle$. However, an explicit pQCD 
calculation\cite{S2} has shown it to be rather small in the energy range 
of interest, i.e. ISR-Tevatron.
Here, I present a re-formulation of the GLM model.
The goal is to construct a multi channel  eikonal model in which the 
rescattering can be either elastic or diffractive.
\par
In the simplest approximation we consider diffraction as a single hadronic 
state. We have, thus, two orthogonal wave functions 
\begin{equation}\label{4.1}
\langle\Psi_{h} \mid \Psi_{d}\rangle\,=\,0.
\end{equation}
$\Psi_{h}$ is the wave function of the incoming hadron, and 
$\Psi_{D}$ is the wave function of the outgoing diffractively produced 
particles, initiated by the incoming hadron.
Denote the interaction operator by {\bf T} and consider two wave functions 
$\Psi_1$ and $\Psi_2$ which are diagonal with respect to {\bf T}.
The amplitude of the interaction is given by
\begin{equation}\label{4.2}
A_{i,k}\,=\,
\langle\Psi_i \Psi_k \mid {\bf T} \mid \Psi_{i^{\prime}} \Psi_{k^{\prime}}\rangle 
\,=\,A_{i,k}^{el}\,\, \delta_{i,i^{\prime}}\,\, \delta_{k,k^{\prime}}.
\end{equation}
In a $2 \times 2$ model $i,k\,=\,1,2$. The amplitude $A_{i,k}^{el}$ 
satisfies the diagonal unitarity condition (see \eq{2.13})
\begin{equation}\label{4.3}
2 Im\,A_{i,k}^{el}(s,b) \,=\, \mid A_{i,k}^{el}(s,b) \mid^2 \,+\,
G_{i,k}^{in}(s,b),
\end{equation}
for which we write the solution
\begin{equation}\label{4.4}
A_{i,k}^{el}(s,b) \,=\, i\(1\,-\,e^{-\frac{\Omega_{i,k}(s,b)}{2}}\),
\end{equation}
and
\begin{equation}\label{4.5}
G_{i,k}^{in} \,=\, 1-e^{-\Omega_{i,k}(s,b)}.
\end{equation}
$\Omega_{i,k}(s,b)$ is the opacity of the $(i,k)-th$ channel 
with a wave function $\Psi_i\, \times \,\Psi_k$.
The probability that two hadronic states $i$ and $k$ do not interact 
inelastically is
\begin{equation}\label{4.5.1}
P_{i,k}(s,b)\,=\,e^{-\Omega_{i,k}(s,b)}\,=\,\{1\,-\,A_{i,k}^{el}(s,b)\}^2.
\end{equation} 
\par
In this representation $\Psi_h$ and $\Psi_d$ can be written as
\begin{equation}\label{4.6}
\Psi_h\,=\,\alpha \Psi_1\,+\,\beta \Psi_2,
\end{equation}
\begin{equation}\label{4.7}
\Psi_d\,=\,-\beta \Psi_1\,+\,\alpha \Psi_2.
\end{equation}
Since $\mid \Psi_h \mid^2\,=\, 1$, we have
\begin{equation}\label{4.8}
\alpha^2\,+\,\beta^2\,=\,1.
\end{equation}
The wave function of the final state is 
\begin{equation}\label{4.9}
\Psi_f\,=\,\mid {\bf T} \mid \Psi_h \times \Psi_h\rangle\,
=\,\alpha^2 A_{1,1} \Psi_1 \times \Psi_1\,+\,
\alpha \beta A_{1,2} \{\Psi_1 \times \Psi_2\,+\,\Psi_2 \times \Psi_1 \}
\,+\,\beta^2 A_{2,2} \Psi_2 \times \Psi_2.
\end{equation}
Since $A_{i,k}^{el}$ is a $2 \times 2$ matrix, we have to consider 4 
possible rescattering processes. However, in the case of a $\bar p p$ 
(or $p p$)
collision, single diffraction at the proton vertex equals single 
diffraction at the antiproton vertex. i.e. $A_{1,2}\,=\,A_{2,1}$ and we 
end with 3 channels whose 
b-space amplitudes are given by 
\begin{equation}\label{4.10}
a_{el}(s,b)\,=\, \langle\Psi_h \times \Psi_h \mid \Psi_f\rangle \,=\,
\alpha^4 A_{1,1} \,+\,
2 \alpha^2 \beta^2 A_{1,2}\,+\,
\beta^4 A_{2,2},
\end{equation}
\begin{equation}\label{4.11}
a_{sd}(s,b)\,=\, \langle\Psi_h \times \Psi_d \mid \Psi_f\rangle \,=\,
\alpha \beta \{\- \alpha^2 A_{1,1}\,+\,(\alpha^2\,-\,\beta^2) A_{1,2}\,+\,
\beta^2 A_{2,2} \},
\end{equation}
\begin{equation}\label{4.12}
a_{dd}(s,b)\,=\, \langle\Psi_d \times \Psi_d \mid \Psi_f\rangle \,=\,
\alpha^2 \beta^2 \{ A_{1,1}\,-\, 2 A_{1,2}\,+\, A_{2,2}\}.
\end{equation}
$A_{i,k}^{el}$ and $G_{i,k}^{in}$ are given by \eq{4.4} and \eq{4.5}. The 
input values of $\Omega_{i,k}$ satisfy the Regge factorization property
\begin{equation}\label{4.13}
\Omega_{2,2}(s,b)\,=\,\frac{\Omega_{1,2}(s,b)\,\Omega_{2,1}(s,b)}
{\Omega_{1,1}(s,b)}.
\end{equation} 
\par
As in the single channel GLM model
we simplify the calculation assuming a Gaussian b-space distribution of
the input opacities $\Omega_{i,k}$.
\begin{equation}\label{4.14.1}
\Omega_{1,1}\,=\,\nu_1 X,
\end{equation}
\begin{equation}\label{4.14.2}
\Omega_{2,2}\,=\,\nu_2 X^{r/(r-2)},
\end{equation}
\begin{equation}\label{4.14.3}
\Omega_{1,2}\,=\,\sqrt{\nu_1 \nu_2 (2\,-\,r)} X^r.
\end{equation}
$X\,=\,e^{-b^2/R_{1,1}^2(s)}$ and $r\,=\,R_{1,1}^2(s)/R_{1,2}^2(s)$.
The factorizable radii are given by
\begin{equation}\label{4.14.4}
R_{i,k}^2\,=\,2 R_{i,0}^2\,+\,2 R_{k,0}^2\,+\,4 \alpha_P^{\prime} 
ln(\frac{s}{s_0}).
\end{equation}
\par
The opacity expressions just presented allow us to express all 
physical observable of interest as functions of $\nu_1, \nu_2$, r and 
$\beta$. The first three variables depend on $s$ while $\beta$ is a 
constant of the model. The determination of these variables 
enables us to produce a global 
fit\cite{GLMMC} to the total 
cross sections as well as the elastic, single and double 
diffractive integrated cross sections. 
This has been done in a two channel 
model, in which $\sigma_{dd}$ is neglected and a more detailed three 
channel model. The main conclusion of these studies is that the 
extension of the GLM model to a multi channel eikonal results with a good 
overall reproduction of the data. The results maintain the  
b-space peripherality of the diffractive output amplitudes and 
satisfies the Pumplin bound.
Note that since different experimental groups 
have been using different algorithms to define diffraction, the SD   
experimental points are too scattered to enable a tight 
theoretical reproduction of the data. 
\begin{figure}
\begin{center}
\psfig{file=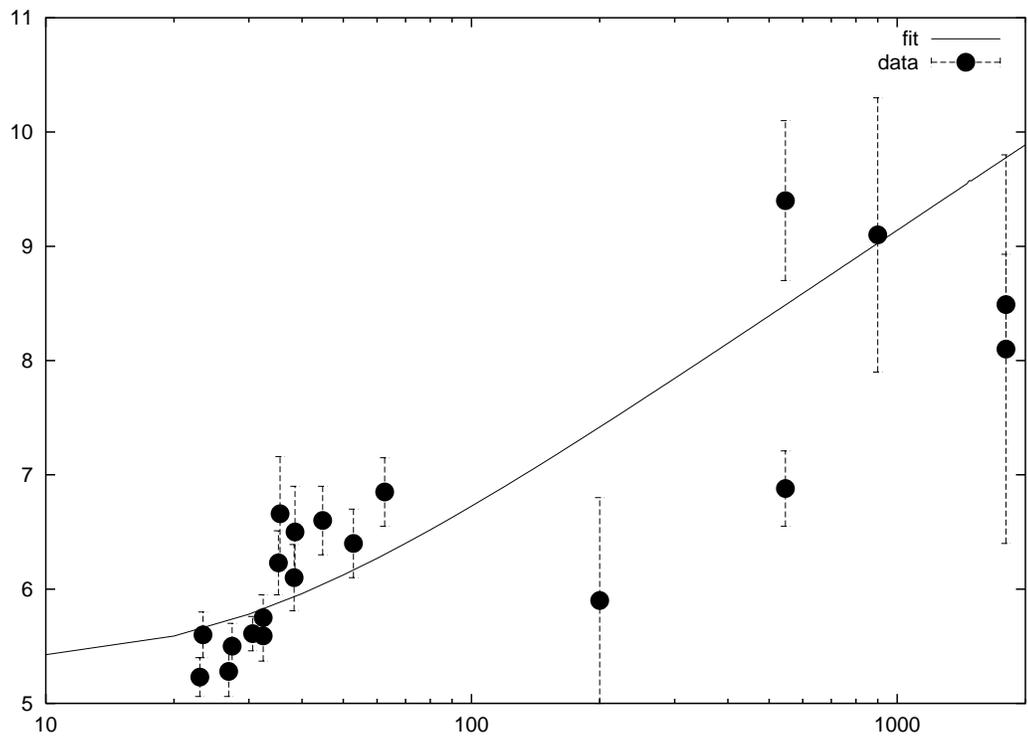,width=12cm,angle=-90}
\end{center}
\caption{\it Integrated SD data and a 3 channel model fit\cite{GLMMC}.}
\end{figure}
For details, see Refs. \cite{Dino},
\cite{GLM} and \cite{GLMMC} and Fig. 6.
\par
To complete this discussion on the generalization of the GLM model we have 
to show that the multi channel model reproduces the excellent 
results we have obtained for the survival probability 
in the simple, single channel, model. This is not self evident.
In the single channel model we have correlated the decrease with energy 
of $\langle\mid S_{spec} \mid^2\rangle$ mainly with the corresponding 
power like increase of 
$R_{el}\,=\,\sigma_{el}/\sigma_{tot}$, see Fig. 7.
\begin{figure}
\begin{center}
\psfig{file=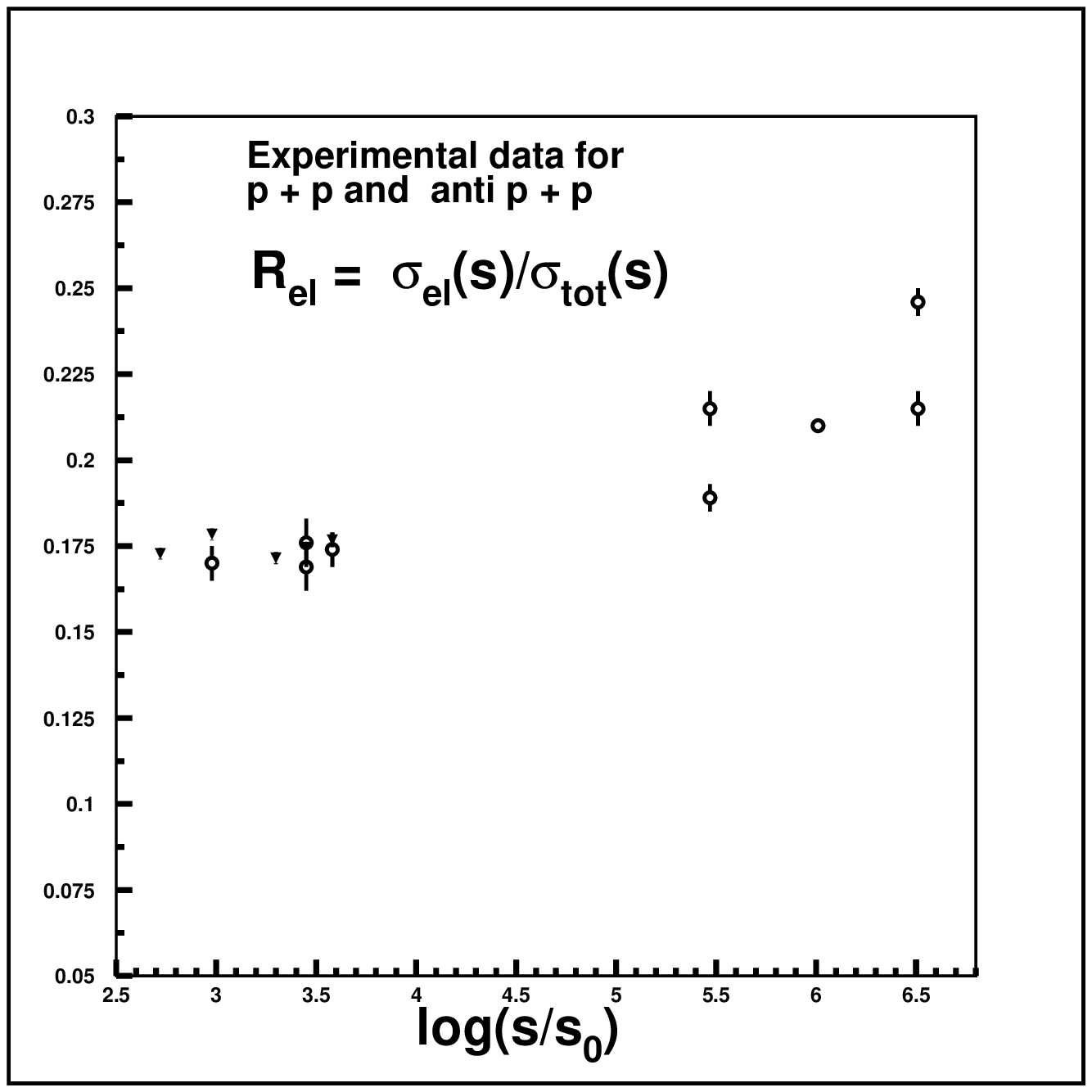,width=12cm,bbllx=100,bblly=300,bburx=510,bbury=710}
\end{center}
\caption{\it The experimental data on the ratio
$R_{el}\,=\,\sigma_{el}/\sigma_{tot}$.}
\end{figure}
In a 3 channel model $R_{el}$ is replaced by 
$R_D\,=\,(\sigma_{el}\,+\,\sigma_{sd}\,+\,\sigma_{dd})/\sigma_{tot}$,
which is approximately constant, see Fig. 8.
\begin{figure}
\begin{center}
\psfig{file=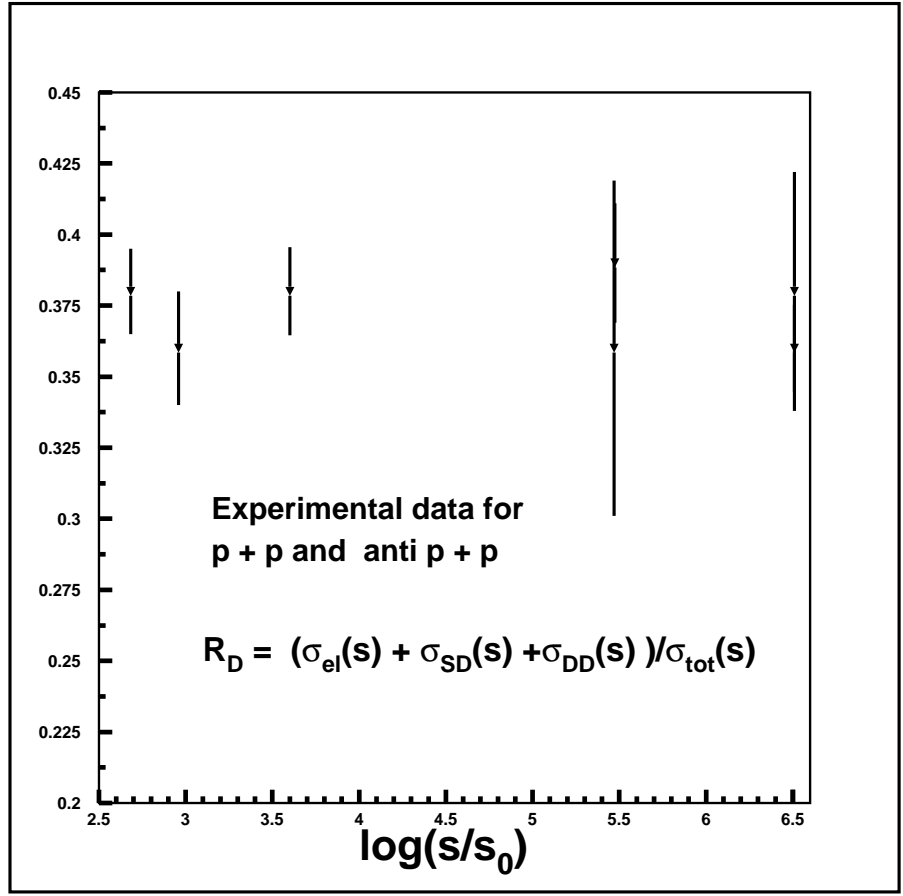,width=12cm,bbllx=100,bblly=300,bburx=510,bbury=710}
\end{center}
\caption{\it The experimental data on the ratio 
$R_D\,=\,(\sigma_{el}\,+\,\sigma_{sd}\,+\,\sigma_{dd})/\sigma_{tot}$.}
\end{figure}
Indeed, a calculation\cite{RR} suggests that 
\begin{equation}\label{RR}
\langle\mid S_{spec} \mid^2\rangle\,=\,\(1\,-\,R_D\)^2,
\end{equation}
which yields a reasonable value of the survival probability, but fails to 
reproduce its energy dependence. Upon investigation, one finds that this 
problem originates 
from the neglect of the differential b-dependence of the various 
scattering amplitudes in Ref. \cite{RR}.
From the 3 channel GLM model we obtain 
\begin{equation}\label{63}
\langle{\mid S_{spec} \mid}^2\rangle\,=\,\frac{N(s)}{D(s)},
\end{equation}
where
\begin{equation}\label{64}
N(s)\,=
\,\int_0 ^1 \frac{dX}{X} 
\{(1\,-\,\beta^2)^2 P_{1,1} \nu_1 a_{1,1} X^{a_{1,1}} 
\,+\,2 (1\,-\,\beta^2) \beta^2 P_{1,2} \nu_{1,2} a_{1,2} X^{a_{1,2} r}
\end{equation}
$$\,+\,\beta^4 P_{2,2} \nu_2 a_{2,2} X^{a_{2,2}r}/(2\,-\,r)\},$$
and 
\begin{equation}\label{65}
D(s)\,=\,\int_0^1 \frac{dX}{X}   
\{(1\,-\,\beta^2)^2 \nu_1 a_{1,1} X^{a_{1,1}}
\,+\,2 (1\,-\,\beta^2) \beta^2 \nu_{1,2} a_{1,2} X^{a_{1,2} r}
\end{equation} 
$$\,+\,\beta^4 \nu_2 a_{2,2} X^{a_{2,2}r}/(2\,-\,r)\}.$$
We denote $\nu_{1,2}\,=\,\sqrt{\nu_1 \nu_2 (2\,-\,r)}$ and 
$a_{i,k}\,=\,\frac{R_{i,k}^2}{{\bar R}_{i,k}^2}$. 
\par
A b-dependant calculation\cite{GLMLRG}\cite{GLMMC} of 
$\langle\mid S_{spec} \mid^2\rangle$ and its energy 
dependence reproduces the D0, CDF and HERA 
results\cite{D0}\cite{CDF}\cite{HERA},  
with $\beta\, \simeq \,0.5$, which is a very reasonable value.
\section{Discussion}
\par
Both the single channel and the improved multichannel GLM models, with 
which we calculate the survival probability and its energy dependence, 
do not attempt to calculate the pQCD ratio $F_s$ which is an external 
input. Accordingly, the GLM model, on its own, cannot provide a 
calculation 
of $f_{gap}$. Never the less, the model provides an important, and 
relatively simple, direct correlation between the soft scattering 
experimental features and 
$\langle{\mid S_{spec} \mid}^2\rangle$. i.e. given the soft scattering 
experimental features and the experimental value of $B_{hard}$, we can 
predict the survival probabilities. As such even the partial experimental 
information we have on $\langle{\mid S_{spec} \mid}^2\rangle$ from LRG 
di-jet production serves as an excellent probe of the roll of s-channel 
unitarity in high energy soft scattering, both elastic and diffractive. 
\par
The GLM model is conceptually different from both the Durham\cite{Durham} 
and the CEM\cite{Halzen} alternative approaches. All three models 
calculate 
$\langle{\mid S_{spec} \mid}^2\rangle$ using the interplay between hard 
and soft physics and obtain similar values for the survival 
probabilities. However, we note that the GLM model does not dependent on 
input free parameters, at the cost of having a more limited predictive 
power than the other two models. 
At the present stage this is not a significant disadvantage  
since the experimental data base is rather small. 
Both Durham and 
CEM are partonic models which have a wider predictive power at the cost of 
depending on a rather extensive input.
Note that even though both GLM and Durham are multichannel models, they
are dynamically different. GLM multi channel calculation relates to the 
diversity of the intermediate rescatterings, i.e. elastic and diffractive. 
In the Durham model the term 
"multi channel" relates to the input information. The model investigates 
two options:
i) Small and large dipoles,
ii) Valence and sea partons.
\newline
Both options give similar results which are compatible with GLM.
\par 
Much attention has been given recently to the compatibility  
between the Tevatron and DESY data. Clearly, rather than depending on 
a pdf input to calculate $F_s$ we may use the gluon structure function 
inferred from the diffractive HERA data\cite{HERA} and use it as input to 
the calculation of $F_s$ at the Tevatron. This is an over 
simplified procedure ignoring the roll of the survival probability,
and it has led to speculations about a possible breaking of QCD 
or Regge factorization or both. 
Once the dependence of 
$\langle{\mid S_{spec} \mid}^2\rangle$ on $s$ and $Q^2$ is investigated 
it is an easy exercise to re establish the compatibility between the 
Tevatron data and DESY photo and DIS diffractive data. For a detailed 
discussion see Refs. \cite{Dino} and \cite{Durham}.
In this context we note that the recent CDF data on multi LRG gap jet 
production\cite{Dino} provides additional support to the philosophy 
advanced in this summary. As long as 
$R_H^2(multi\,LRG)\,\simeq\,R_H(LRG)$
we anticipate the two survival probabilities to be approximately the same 
as, indeed, is reported experimentally.
\newline
\newline
\newline
\section {Acknowledgements}
Most of the physics presented in this review is a product of a long 
standing collaboration with E. Gotsman and E. Levin. I wish to thank both 
of them for being such good colleagues and friends. Much of the writing of 
this paper was done when I was visiting UERJ. It is a pleasure to thank 
A. Santoro and his group for their hospitality and for creating such a 
stimulating island of physics studies in the heart of Rio. 
\newpage
\newcommand{\refbrake}{\\\hspace*{2mm}}


\begin{thebibliography}{99}
\bibitem{Dok} 
Yu.L. Dokshitzer, V. Khoze and S.I. Troyan:
{\it Sov. J. Nucl. Phys.} {\bf 46} (1987) 712.\\
Yu.L. Dokshitzer, V. Khoze and T. Sjostrand: \plb{274}{1992}{116}.
\bibitem{Bj}
J. D. Bjorken:
{\it Int. J. Mod. Phys.} {\bf A7} (1992) 4189;
\prd{47}{1993}{101}.
\bibitem{D0}
D0 Collaboration:
\prl{72}{1994}{2332}; \prl{76}{1994}{734}; \plb{440}{1998}{189}.
\bibitem{CDF}
CDF Collaboration:
\prl{74}{1995}{855}; \prl{80}{1998}{1156}; \prl{81}{1998}{5278};
\prl{84}{2000}{5043}; \prl{85}{2000}{4215}.
\bibitem{Dino}
K. Goulianos:
{\it Proceedings of Diffraction 2002, Alushta (Crimea)}, 
Kluwer Academic Pub. (2002) 13; {\it J. Phys.} {\bf G26} (2000) 716; 
{\it Nucl. Phys. Proc. Suppl.} {99A} (2001) 37.\\ 
See also: 
hep-ph/0203141 and hep-ph/0205217. 
\bibitem{HERA}
ZEUS Collaboration:
\plb{315}{1993}{481}; \zpc{68}{1995}{569}; \plb{369}{1996}{55}.\\
H1 Collaboration: \npb{429}{1994}{477}.\\
A.A. Savin: 
{\it Proceedings of Diffraction 2002, Alushta (Crimea)},
Kluwer Academic Pub. (2002) 23.
\bibitem{GLMLRG}
E. Gotsman, E.M. Levin and U. Maor:
\plb{309}{1993}{199}; \npb{493}{1997}{354}; \plb{438}{1998}{229};
\prd{60}{1999}{094011}.
\bibitem{GLM}
E. Gotsman, E.M. Levin and U. Maor:
\zpc{57}{1993}{667}; \prd{49}{1994}{R4321}; \plb{353}{1995}{526};
\plb{347}{1995}{424}.
\bibitem{Durham}
V.A. Khoze, A.D. Martin and M.G. Ryskin:
\epj{C14}{2000}{525}; \epj{C18}{2000}{167}; \epj{C19}{2001}{477}.\\
A.B. Kaidalov, V.A. Khoze, A.D. Martin and M.G. Ryskin:
\epj{C21}{2001}{521}.
\bibitem{Halzen}
O.J.P. Eboli, E.M.Gregores and F. Halzen:
\prd{61}{2000}{034003}; {\it Nucl. Phys. Proc. Suppl.} {\bf 99A} (2001) 
257.\\ 
M.M. Block and F. Halzen: \prd{63}{2001}{114004}.
\bibitem{S2}
V.A. Khoze, A.D. Martin and M.G. Ryskin:
\plb{401}{1997}{330}; \prd{56}{1997}{5867}.\\ 
G. Oderda and G. Sterman: \prl{81}{1998}{3591}.
\bibitem{Regge}
P.D.B. Collins:
{\it An Introduction to Regge Theory and High Energy Physics}, 
Cambridge University Press (1977).\\
L. Caneschi (editor):  
{\it Current Physics Sources and Comments Vol. 3: 
Regge Theory of Low $P_T$ Hadronic Interactions}, 
North Holland Pub (1989). 
\bibitem{DL}
A. Donnachie and P.V. Landshoff:
\npb{231}{1984}{189}; \plb{296}{1992}{227}; \zpc{61}{1994}{139}.
\bibitem{BKW}
M.M. Block, K. Kang and A.R. White:
{\it Mod. Phys.} {\bf A7} (1992) 4449.
\bibitem{Mueller}
A.H. Mueller:
\prd{2}{1970}{2963}; \prd{4}{1971}{150}.
\bibitem{Froissart}
M. Froissart: 
{\it Phys. Rev.} {\bf 123} (1961) 5535.                                      
\bibitem{CDFb}
CDF Collaboration: 
\prd{50}{1994}{5535}.
\bibitem{Pumplin}
J.D. Pumplin: 
\prd{8}{1973}{2849}; {\it Physica Scripta} {25A} (1982) 191.
\bibitem{BSW}
C. Bourrely, J. Soffer and T.T. Wu:
\plb{252}{1990}{287}; \plb{339}{1994}{322}.
\bibitem{HERAJ}
H1 Collaboration: 
\plb{483}{2000}{23}.\\
ZEUS Collaboration: 
\epj{C24}{2002}{345}. 
\bibitem{CDF2p}
CDF Collaboration: 
{\it FERMILAB preprint} {\bf Pub-97}, 083-E.
\bibitem{GLMMC}
E. Gotsman, E. Levin and U. Maor:
\plb{452}{1999}{387}.\\
Tz. Gutman: 
{\it TAU M.Sc. Thesis, unpublished} (1993).
\bibitem{RR}
A. Rostovtsev and M.G. Ryskin:
\plb{390}{1997}{375}.

\end{thebibliography}
\end{document}